\documentclass[12pt,a4paper]{article}

\usepackage{SGpreprint}
\usepackage{url,lastpage}

\nologohead{Hans-Ulrich Stark, Claudio J. Tessone,
  Frank Schweitzer\\
  Decelerating microdynamics can accelerate macrodynamics in the voter model
\\  \textsl{Physical Review Letters} \textbf{101} (2008) 018701 \\ 
See \url{http://www.sg.ethz.ch} for
more info \mbox{}
}

\usepackage{epsfig}
\usepackage{amsmath}

\newcommand{\abs}[1]{\left| #1 \right|}

\begin{document}

\begin{center}

  \textbf{\Large Decelerating microdynamics can accelerate macrodynamics in
    the voter     model}\\[5mm]

  \textbf{Hans-Ulrich Stark,
Claudio J. Tessone, 
Frank Schweitzer}

 {Chair of Systems Design, ETH Zurich, Kreuzplatz 5, 8032
      Zurich, Switzerland}
\end{center}

\begin{abstract}
  For the voter model, we study the effect of a memory-dependent
  transition rate. We assume that the transition of a spin into the
  opposite state decreases with the time it has been in its current
  state.  Counter-intuitively, we find that the time to reach a
  macroscopically ordered state can be accelerated by slowing-down the
  microscopic dynamics in this way.  This holds for different network
  topologies, including fully-connected ones. We find that the ordering
  dynamics is governed by two competing processes which either stabilize
  the majority or the minority state. If the first one dominates, it
  accelerates the ordering of the system. The conclusions of this Letter
  are not restricted to the voter model, but remain valid to many other
  spin systems as well.


\end{abstract}

How fast an out-of-equilibrium system reaches an ordered state has been a
central question in statistical physics, but also also in disciplines
such as chemistry, biology, and social sciences.  Despite its simple
structure, the ``voter model'' has served as a paradigm to study this
question \citep{ligget1995}. It is one of the few spin systems that can
be analytically solved in regular lattices \cite{redner2001}.  In
physics, it was investigated how the time to reach the equilibrium state
depends on the system size, the initial configuration, and the topology
of the interactions~\citep{col1}.
Among its prominent properties, the magnetization conservation has been
studied extensively~\citep{col2}.
Furthermore, the formation and growth of state domains was studied,
showing the existence of coarsening without surface tension in
two-dimensional systems~\citep{col3}.
The voter model also found numerous interdisciplinary applications,
e.g.~in chemical kinetics \citep{krapivsky1992} and in
ecological~\citep{ravasz04,col4}
and social systems~\citep{galam2005}.  Its properties have also served to
complete the understanding of other spin-systems, such as the Ising model
and spin-glasses~\cite{col5}.

To assume that transition rates are constant in time is (in general) not
valid for non-equilibrium systems. A good example are spin glasses, where
the effective temperature of the system changes with the time elapsed
since a given perturbation was applied~\cite{cugliandolo1993}.  In this
Letter, we consider that, for each site, the transition rates are not
constant, but {\em decrease} with the time elapsed since the last change
of state (namely, its {\em persistence time}). We refer to this change as
{\em increasing inertia}. The level of inertia is measured by how fast
the transition rates decrease with persistence time.  Dependent on the
context of the voter model, this mechanism has different interpretations.
In a social context, the longer a voter already stays with its current
state, the less it may be inclined to change it in the next time step,
which can be interpreted as conviction. In models of species
competition~\cite{ravasz04}, this would imply that neighboring species
are less likely to be displaced at a later stage of growth.

Obviously, increasing inertia leads to slower microscopic dynamics.
Against intuition and in contrast to results with fixed (homogeneous or
heterogeneous) values of inertia, we find
that 
the time to reach an ordered state can be effectively reduced. We further
find that this phenomenon exists independently of the exact network
topology in which the system is embedded.  We show that the unexpected
reduction of the time to reach an ordered state is related to the break
of magnetization conservation, which holds for the standard voter model.
This break originates from the evolving heterogeneity in the transition
probabilities within the voter population, which, in the extended model,
depends on the distribution of the persistence times.

The voter model denotes a simple binary system comprised of $N$ {\em
  voters}, each of which can be in one of two states (often referred to
as {\em opinions}), $\sigma_{i}= \pm 1$.  
A voter is selected at random and adopts the state of a randomly chosen
neighbor. After $N$ such update events, time is increased by 1. In this
work, we consider homogeneous networks, where all voters have the same
number of neighbors.  In the standard voter model, the transition rate at
which voter $i$ switches to the opposite state, $\omega^{V}(-\sigma_i |
\sigma_i)$, is proportional to the frequency of state $-\sigma_i$ in $
\lbrace i \rbrace$, the set of the $k$ neighbors of $i$, namely
\begin{equation}
  \omega^{V}(-\sigma_i | \sigma_i) = \frac{\beta}{2}
  \left(1-\frac{\sigma_i}{k} \sum_{j\in \lbrace  i \rbrace  }\sigma_j\right).
  \label{lvm}
\end{equation}
The prefactor $\beta$ determines the time scale of the transitions and is
set to $\beta=1$.  In order to describe the dynamics on the macrolevel,
we introduce the global densities of voters with state $+1$ as $A(t)$ and
with state $-1$ as $B(t)$. The instantaneous magnetization is then given
by $M(t) = A(t) - B(t)$.  Starting from a random distribution of states,
we have $M(0)=0$. The emergence of a completely ordered state (which is often
referred to as {\em consensus}), is characterized by $\abs{M}=1$. The time to
to reach consensus, $T_{\kappa}$, is obtained through
an average over many realizations. The dynamics of the global frequencies is
formally given by
the rate equation
$$
\dot A(t) = - \dot B(t) =  \Omega^V(+1|-1) B(t) - \Omega^V(-1|+1) A(t).
$$ 
The \emph{macroscopic} transition rates $ \Omega^V$ have to be obtained
from the aggregation of the microscopic dynamics given by Eq.~(\ref{lvm}). A
simple expression for these can be found in the mean-field
limit. There, it is assumed that the frequencies of states in the
local neighborhood can be replaced by the global ones. This gives
$\Omega^V(+1|-1) = A(t)$,
$\Omega^V(-1|+1) = B(t)$ and leads to
$\dot A(t) = A(t)B(t) -  B(t)A(t)\equiv 0$.
For an ensemble average, the frequency of the outcome of a particular
consensus state $+1$ is equal to the initial frequency $A(0)$ of state
$+1$, which implies the conservation of magnetization.  It is worth
noticing that, for a single realization, the dynamics of the voter model
is a fluctuation driven process that, for finite system sizes, always
reaches consensus towards either $+1$ or $-1$.  We now investigate how
this dynamics changes if we modify the voter model by assuming that
voters additionally have an inertia $\nu_i \in [0,1]$ which leads to a
decrease of the transition rate to change their state
\begin{equation}
  \label{ivm}
  \omega(-\sigma_i|\sigma_i, \nu_i)=(1-\nu_i)\,\omega^V(-\sigma_i| 
  \sigma_i) .
\end{equation}
Obviously, if all voters have the same fixed value of inertia
$\nu_\bullet$, the dynamics is equivalent to the standard voter model
with the time scaled by a factor $(1-\nu_\bullet)^{-1}$. Similar results
are obtained if the inertia values are randomly distributed in the
system: higher consensus times are found for increasing levels of
inertia. In our model, however, we consider an individual and evolving
inertia $\nu_{i}$ that depends on the persistence time $\tau_i$ the voter
has been keeping its current state.  For the sake of simplicity, the
results presented here assume that the individual inertia $\nu_{i}$
increases linearly with persistence time $\tau_{i}$, $\mu$ being the
``strength'' of this response, until it reaches a saturation value
$\nu_s$, i.e.
 $\nu(\tau_i) = \mathrm{min}\left[\mu\,\tau_i ,\nu_s \right]$.
Choosing $\nu_s < 1$ avoids trivial frozen states of the dynamics
\footnote{The results presented here are qualitatively independent of the
  exact functional relation $\nu_i(\tau_i)$, as long as a monotonously
  increasing function with a saturation below 1 is considered.}. The rate
of inertia growth $\mu$ determines the number of timesteps until the
maximal inertia value is reached, denoted as
${\tau_s}=\left[{\nu_s}/{\mu}\right]$. 

\begin{figure*}[ht!]
  \begin{centering}
    \includegraphics[width=0.95\linewidth]{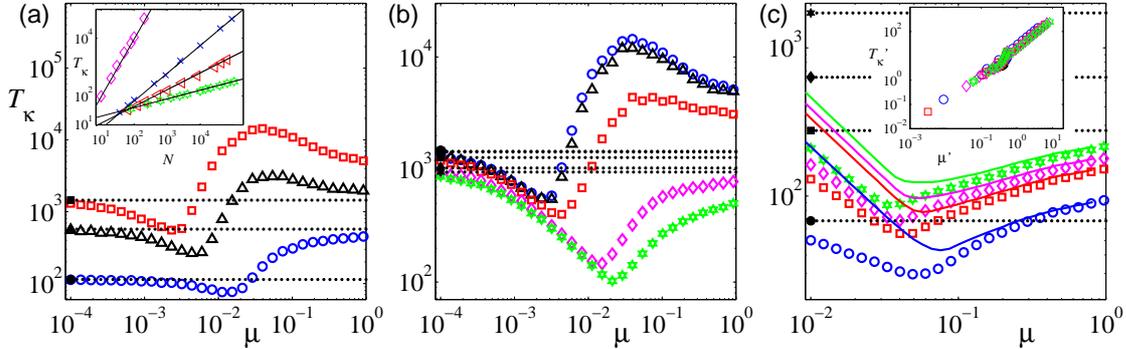}
  \end{centering}
\vspace*{-0.5cm}
  \caption{\label{ttc}(color online). Average consensus times $T_\kappa$
    for varying values of the inertia slope $\mu$ and fixed saturation
    value $\nu_{s}=0.9$. Sample sizes vary between $10^3-10^4$ simulation
    runs. Filled, black symbols always indicate the values of $T_\kappa$
    at $\mu=0$.  \textbf{(a)} $2d$ regular lattices ($k_{i}=4$) with
    system sizes: ($\circ$) $N=100$, ($\triangle$) $N=400$, ($\square$)
    $N=900$. The inset shows how consensus time scales with
      system size in regular lattices at $\mu = \mu^*$: ($\diamond$) $1d$,
($\times$)
      $2d$, ($\triangleleft$) $3d$, ($\star$) $4d$. \textbf{(b)}
    Small-world networks obtained by randomly rewiring a $2d$ regular
    lattice with probability: ($\circ$) $p_r=0$, ($\triangle$)
    $p_r=0.001$, ($\square$) $p_r=0.01$, ($\diamond$) $p_r=0.1$,
    ($\star$) $p_r=1$.  The system size is $N=900$.  \textbf{(c)} Fully
    connected networks (mean field case, $k_{i}=N-1$) with system sizes:
    ($\circ$) $N=100$, ($\square$) $N=900$, ($\diamond$) $N=2500$,
    ($\star$) $N=10^4$. Lines represent the numerical solutions of
    Eqs.~(\ref{eq:dyn:atau}), (\ref{eq:dyn:a0}), (\ref{eq:I}) with the
    specifications in the text. The inset shows the collapse of the
    simulation curves by scaling $\mu$ and $T_\kappa$ as explained in the
    text. }
\end{figure*}

Increasing $\mu$ increases the level of inertia within the voter
population, thereby slowing-down the microscopic dynamics. Like in the
case with fixed inertia, one would intuitively assume an increase of the
average time to reach consensus.  Interestingly, this is not always the
case as simulation results of $T_{\kappa}(\mu)$ show for different
network topologies (see Fig.~\ref{ttc}). Instead, it is found that there
is an intermediate value $\mu^*$, which leads to a global minimum in
$T_\kappa$~\footnote{In this Letter, we do not investigate the origin of
  the global maxima in the consensus times of Fig.~\ref{ttc} (a). In
  contrast to the global minima, this effect results from spatial
  configurations as can be learned from panels (b) and (c) of the same
  figure.}.  For $\mu <\mu^\ast$, consensus times decrease with
increasing $\mu$ values.  Only for $\mu >\mu^\ast$, higher levels of
inertia result in increasing consensus times.

For a two-dimensional lattice, shown in Fig.~\ref{ttc}(a), we find $\mu^*
\propto 1/\ln N$. Simulations of regular lattices in other dimensions
show that the non-monotonous effect on the consensus times is amplified
in higher dimensionality of the system. Being barely noticeable for
$d=1$, the ratio between $T_\kappa(\mu^\ast)$ and $T_\kappa(\mu=0)$
(i.e.~the standard voter model) decreases for $d=3$ and $d=4$. We further
compare the scaling of $T_{\kappa}$ with system size $N$ for the standard
and the modified voter model. The first one gives for one-dimensional
regular lattices ($d=1$) $T_\kappa\propto N^2$ and for two-dimensional
regular lattices ($d=2$) $T_\kappa\propto N \log N$. For $d>2$ the system
does not always reach an ordered state in the thermodynamic limit. In
finite systems, however, one finds $T_{\kappa} \sim N$.  In the modified
voter model, we instead find that $T_{\kappa}(\mu^\ast)$ scales with
system size as a power-law, $T_\kappa(\mu^\ast)\propto N^\alpha$ (see
inset in Fig.~\ref{ttc}(a)); where $\alpha=1.99\pm 0.14$ for $d=1$ (i.e.,
in agreement with the standard voter model); $\alpha=0.98\pm 0.04$ for
$d=2$; $\alpha=0.5\pm 0.08$ for $d=3$; and $\alpha=0.3\pm 0.03$ for
$d=4$. For fixed values of $\mu>\mu^*$, the same scalings apply.

In order to cope with the network topology, in Fig.~\ref{ttc}(b) we plot
the dependence of the consensus times $T_\kappa$ for small-world
networks 
built with different rewiring probabilities.  The degree of each node is
kept constant by randomly selecting a pair of edges and exchanging their
ends with probability $p$ \citep{maslov2003}.  It can be seen that the
effect of reduced consensus times for intermediate values of $\mu$ still
exists and is amplified by increasing the randomness of the network. This
result implies that the spatial extension of the system, e.g. in regular
lattices, does not play a crucial role in the emergence of this
phenomenon.  This can be confirmed by investigating the case shown in
Fig.~\ref{ttc}(c), in which the neighborhood network is a fully-connected
one (the solid lines correspond to a theoretical approximation introduced
below). The inset shows the results of a scaling analysis, exhibiting the
collapse of all the curves by applying the scaling relations $\mu' =
\left| \mu \ln(\eta\,N) - \mu_1 \right|$, and $T_\kappa' = T_\kappa /
\ln(N/\xi) \mu'$, with $\eta = 1.8(1)$, $\mu_1 = 1.5(1)$, $\xi = 7.5(1)$.
This shows that the location of the minimum, as well as $T_\kappa$,
scales logarithmically with $N$.

The fact of reaching a final state faster by decelerating the dynamics
microscopically has some resemblances with the ``slower-is-faster''
effect discovered in panic research~\cite{helbing}. However, the origin
of the phenomenon discussed here is quite different, as we can
demonstrate by the following analytical approach. First, note that voters
are fully characterized by their current state $\pm 1$ and their
persistence time $\tau$. Thus, we introduce the global frequencies
$a_\tau(t)$, $b_\tau(t)$ for subpopulations of voters with state $+1$,
$-1$ (respectively) and persistence time $\tau$.  Thus, these frequencies
satisfy
\begin{equation}
  A(t)=\sum_{\tau} a_\tau(t), \;
  B(t)=\sum_{\tau} b_\tau(t).\label{sums}
\end{equation}
Formally, the rate equations for the evolution of these subpopulations in 
the mean-field limit are given by
\begin{eqnarray}
\dot{a}_\tau(t)  = & &\sum_{\tau' } 
\Big[ \Omega( a_\tau | a_{\tau'} ) a_{\tau'} + \Omega( a_\tau | b_{\tau'} ) 
b_{\tau'}  \Big] \nonumber \\
& -&  \sum_{\tau' } \Big[ \Omega( a_{\tau'} | a_\tau ) + \Omega(
b_{\tau'} | a_\tau)
\Big] a_\tau.  \label{eq:atau-rate}
\end{eqnarray}
Due to symmetry, the expressions for $\dot b_\tau(t)$ are
obtained by consistently exchanging $A\leftrightarrow B$ and
$a_\tau\leftrightarrow b_\tau$.

Note that most of the terms in Eq.~(\ref{eq:atau-rate}) vanish because
for a voter only two transitions are possible: (i) it changes its state,
thereby resetting its $\tau$ to zero, or (ii) it keeps its current state
and increases its persistence time by one.  Case (i) is associated with
the transition rate $\Omega(b_0|a_{\tau})$, that in the mean-field limit
reads $\Omega(b_0|a_{\tau })=(1-\nu(\tau))\,B(t)$.  $B(t)$ is the
frequency of voters with the opposite state that trigger this transition,
while the prefactor $(1-\nu(\tau))$ is due to the inertia of voters of
class $a_{\tau}$ to change their state.  For case (ii),
$\Omega(a_{\tau+1}|a_{\tau })=1 - \Omega(b_0|a_{\tau })$, since no voter
can remain in the same subpopulation. I.e., in the mean-field limit, the
corresponding transition rates are $\Omega(a_{\tau+1}|a_{\tau})=
A(t)+\nu(\tau) B(t)$. Therefore, if $\tau>0$, Eq.~(\ref{eq:atau-rate})
reduces to
\begin{eqnarray}
\dot{a}_\tau(t) &=& \Omega(a_{\tau}|a_{\tau-1})\, a_{\tau-1}(t)
- a_\tau(t) \nonumber \\
&=&\Big[ A(t)+ \nu (\tau-1) B(t)\Big] a_{\tau-1}(t) -
  a_\tau(t).
\label{eq:dyn:atau}
\end{eqnarray}
On the other hand, voters with $\tau=0$ evolve as
\begin{eqnarray}
\dot{a}_0(t) &=& \sum_{\tau} \Omega_{b}(a_{0}|b_{\tau}) b_{\tau}(t)
- a_{0}(t)  \nonumber \\
&=& A(t)\Big[ B(t) - I_B(t) \Big] - a_0(t). \label{eq:dyn:a0}
\end{eqnarray}
Due to the linear dependence of the transition rates on inertia, the
terms involving $\nu$ can be comprised into $I_B(t)$ and $I_A(t)$, namely
the average inertia of voters with state $-1$ and $+1$, respectively,
i.e.
\begin{eqnarray}
I_A(t) = \sum_{\tau} \nu( \tau ) a_\tau(t) \,;\,\,\,\,
I_B(t) = \sum_{\tau} \nu( \tau ) b_\tau(t).
\label{eq:I}
\end{eqnarray}
Expressions (\ref{eq:dyn:atau}, \ref{eq:dyn:a0}, \ref{eq:I}) and the
corresponding ones for subpopulations $b_{\tau}$ can be used to give an
estimate of the time to reach consensus in the mean-field limit. Let us
consider an initial state $a_0(t) = A(0) = 1/2 + N^{-1}$ and $b_0(t) =
B(0) = 1/2 - N^{-1}$, i.e. voters with state $+1$ are in slight majority.
By neglecting fluctuations in the frequencies (which drive the dynamics
in the standard voter model), these equations are iterated until
$B(t')<N^{-1}$ (i.e. for a system size $N$, if the frequency of the
minority state falls below $N^{-1}$, the absorbing state is reached).
Then, we assume $T_\kappa = t'$.  The full lines in Fig.~\ref{ttc}(c)
show the results of this theoretical approach, exhibiting the minimum and
displaying good agreement with the simulation results for large values of
$\mu$. For low values of $\mu$, fluctuations drive the system {\em
  faster} into consensus compared to the deterministic approach.

Inserting Eqs.~(\ref{eq:dyn:atau}, \ref{eq:dyn:a0}) into the time-derivative
of Eq.~(\ref{sums}) yields, after some straightforward algebra, the time
evolution of the global frequencies
\begin{eqnarray}
  \dot A(t)=I_A(t)\, B(t)-I_B(t)\, A(t).
    \label{evol-general}
\end{eqnarray}
Remarkably, the magnetization conservation is now broken because of the
influence of the evolving inertia in the two possible states. For
$\nu(\tau) = \nu_\bullet$ (that includes the standard voter model,
$\nu_\bullet=0$), we regain the magnetization conservation. Interestingly
enough, Eq.~(\ref{evol-general}) implies that the frequency $A(t)$ grows
iff.~$I_{A}(t) / A(t) > I_{B}(t) / B(t)$.

When the time dependence of the inertia on the persistence time is a
linear one, as assumed in this Letter, inserting Eqs.~(\ref{eq:dyn:atau},
\ref{eq:dyn:a0}) into Eq.~(\ref{eq:I}) we obtain an equation for the time
evolution of $I_A(t)$ up to first order in $\mu$: 
\begin{eqnarray}
  \dot I_A (t) = A(t)\,I_A(t) + \mu A^{2}(t) - I_A(t) + {\mathcal O}(\mu^2,a_T).
 \label{eq:I:smallmu}
\end{eqnarray}
Here, $a_T=\sum_{\tau\ge\tau_s}a_\tau$ contains all subpopulations with maximum
inertia. Eqs.~(\ref{evol-general}, \ref{eq:I:smallmu}) correspond to a
macroscopic level description of this model. This system of equations has
a saddle point, $A=B=1/2$, $I_A=I_B=\mu /2 + {\mathcal O}(\mu^2)$, and
two stable fixed points, one at $A=1,\ I_A=\nu_s$ and another at $B=1,\
I_B=\nu_s$.  Note that the saddle point is close to the initial condition
of the simulations. Neglecting fluctuations, the time to reach consensus
has two main contributions: (i) the time to escape from the saddle point,
$T_s$; and (ii) the time to reach the stable fixed point, $T_f$; namely
$T_\kappa \sim T_s + T_f$. We then linearize the system around the fixed
points and calculate the largest eigenvalues $\lambda_s$ and $\lambda_f$
(for the saddle and the stable fixed points, respectively) as a function
of $\mu$. A simple argument shows that $T_{s,f} \sim \ln N /
|\lambda_{s,f} (\mu)|$.  At the saddle point, we find $\lambda_s(\mu) =
\sqrt{1+20 \mu + 4 \mu^2}-2\mu-1 + \mathcal{O}(\mu^2)$, which equals to 0
at $\mu=0$ and monotonously increases with $\mu$. For larger values of
$\mu$, where the first order term expansion is no longer valid, numerical
computations show that $\lambda_s$ continues to increase monotonously
with $\mu$.  This means that for larger inertia growth rates $\mu$, the
system will escape faster from the saddle point, thereby reducing the
contribution $T_s$ to the consensus time $T_\kappa$. On the other hand,
for $\mu\to 0$, $\lambda_s$ vanishes and the system leaves the saddle
point only due to fluctuations.

Near the stable fixed points the contribution of $a_T$ to
Eq.~(\ref{eq:I:smallmu}) cannot be neglected anymore. We then obtain
$\lambda_{f,1} = -\nu_s$ for $\mu<1-\nu_s$, whilst $\lambda_{f,2} = \mu -
1$ for $\mu \geq 1-\nu_s$.  Interestingly, both reflect different
processes: the eigenvalue $\lambda_{f,1}$ is connected to voters sharing
the majority state which are, at the level of $\nu_s$, inertial to adopt
the minority one (signalled by $\lambda_{f,1}$ being constant).  For
$\mu\geq 1-\nu_s$, the largest eigenvalue $\lambda_{f,2}$ is related to
voters with the minority state that are, for increasing $\mu$, more
inertial to adopt the majority state (apparent by the decrease in
$|\lambda_{f,2}|$).

The contributions $T_s$ and $T_{f}$ are two competing factors in the
dynamics towards consensus. Qualitatively, they can be understood as
follows: in the beginning of the dynamics, the inertia mechanism
amplifies any small asymmetry in the initial conditions. While this
causes faster time to consensus for (small) increasing values of $\mu$,
for sufficiently large values of inertia growth, another process
outweighs the former: the rate of minority voters converting to the final
consensus state is considerably reduced, too. It is worth mentioning that
the phenomenon described here is robust against changes in the initial
condition: starting from $I_A=I_B<\nu_s$, it holds for any initial
frequencies of opinions. Conversely, starting from $A=B=1/2$, it holds
for any $I_A\ne I_B$.

Summarizing, we investigated the role of microscopically time-dependent
transition rates. In particular, we consider that the microscopic
transition rates decrease with the time elapsed since the last state
change of a given site (called {\em inertia}). Counterintuitively, we
find that intermediate inertia values may lead to much lower times to
reach the absorbing state, i.e.~an accelerated dynamics. It is important
to emphasize that this final state is not an arbitrary one, but most
interestingly, it is always the {\em ordered} one. The mechanism behind
this phenomenon is the existence of two competing processes near the
initial condition and absorbing states. Due to the general analytical
approach taken in this Letter, we emphasize that this phenomenon is not
restricted to the voter model, but is expected to appear near the
absorbing states of any spin system, whenever the inertia mechanism is
present.

{\bf Acknowledgment} 
CJT acknowledges financial support by SBF (Switzerland) through research
project C05.0148 (Physics of Risk).


\begin{thebibliography}{22}
\expandafter\ifx\csname natexlab\endcsname\relax\def\natexlab#1{#1}\fi
\expandafter\ifx\csname bibnamefont\endcsname\relax
  \def\bibnamefont#1{#1}\fi
\expandafter\ifx\csname bibfnamefont\endcsname\relax
  \def\bibfnamefont#1{#1}\fi
\expandafter\ifx\csname citenamefont\endcsname\relax
  \def\citenamefont#1{#1}\fi
\expandafter\ifx\csname url\endcsname\relax
  \def\url#1{\texttt{#1}}\fi
\expandafter\ifx\csname urlprefix\endcsname\relax\def\urlprefix{URL }\fi
\providecommand{\bibinfo}[2]{#2}
\providecommand{\eprint}[2][]{\url{#2}}

\bibitem[{\citenamefont{Liggett}(1995)}]{ligget1995}
\bibinfo{author}{\bibfnamefont{T.~M.} \bibnamefont{Liggett}},
  \emph{\bibinfo{title}{Interacting Particle Systems}}
  (\bibinfo{publisher}{Springer}, \bibinfo{address}{New York},
  \bibinfo{year}{1995}).

\bibitem[{\citenamefont{Redner}(2001)}]{redner2001}
\bibinfo{author}{\bibfnamefont{S.}~\bibnamefont{Redner}},
  \emph{\bibinfo{title}{A guide to first-passage processes}}
  (\bibinfo{publisher}{Cambridge University Press},
  \bibinfo{address}{Cambridge}, \bibinfo{year}{2001}).

\bibitem[{\citenamefont{Sood and Redner}(2005)}]{col1}
\bibinfo{author}{\bibfnamefont{V.}~\bibnamefont{Sood}} \bibnamefont{and}
  \bibinfo{author}{\bibfnamefont{S.}~\bibnamefont{Redner}},
  \bibinfo{journal}{Phys. Rev. Lett.} \textbf{\bibinfo{volume}{94}},
  \bibinfo{eid}{178701} (\bibinfo{year}{2005});
\bibinfo{author}{\bibfnamefont{C.}~\bibnamefont{Castellano}},
  \bibinfo{author}{\bibfnamefont{V.}~\bibnamefont{Loreto}},
  \bibinfo{author}{\bibfnamefont{A.}~\bibnamefont{Barrat}},
  \bibinfo{author}{\bibfnamefont{F.}~\bibnamefont{Cecconi}}, \bibnamefont{and}
  \bibinfo{author}{\bibfnamefont{D.}~\bibnamefont{Parisi}},
  \bibinfo{journal}{Phys. Rev. E} \textbf{\bibinfo{volume}{71}},
  \bibinfo{eid}{066107} (\bibinfo{year}{2005});
\bibinfo{author}{\bibfnamefont{M.}~\bibnamefont{Mobilia}},
  \bibinfo{journal}{Phys. Rev. Lett.} \textbf{\bibinfo{volume}{91}},
  \bibinfo{pages}{028701} (\bibinfo{year}{2003});
\bibinfo{author}{\bibfnamefont{K.}~\bibnamefont{Suchecki}},
  \bibinfo{author}{\bibfnamefont{V.~M.} \bibnamefont{Egu{\'i}luz}},
  \bibnamefont{and}
  \bibinfo{author}{\bibfnamefont{M.}~\bibnamefont{San~Miguel}},
  \bibinfo{journal}{Phys. Rev. E} \textbf{\bibinfo{volume}{72}},
  \bibinfo{pages}{036132} (\bibinfo{year}{2005}{\natexlab{a}}).

\bibitem[{\citenamefont{Frachebourg and Krapivsky}(1996)}]{col2}
\bibinfo{author}{\bibfnamefont{L.}~\bibnamefont{Frachebourg}} \bibnamefont{and}
  \bibinfo{author}{\bibfnamefont{P.}~\bibnamefont{Krapivsky}},
  \bibinfo{journal}{Phys. Rev. E} \textbf{\bibinfo{volume}{53}},
  \bibinfo{pages}{R3009} (\bibinfo{year}{1996});
\bibinfo{author}{\bibfnamefont{F.}~\bibnamefont{Slanina}} \bibnamefont{and}
  \bibinfo{author}{\bibfnamefont{H.}~\bibnamefont{Lavicka}},
  \bibinfo{journal}{Eur. Phys. J. B} \textbf{\bibinfo{volume}{35}},
  \bibinfo{pages}{279} (\bibinfo{year}{2003});
\bibinfo{author}{\bibfnamefont{C.}~\bibnamefont{Castellano}},
  \bibinfo{author}{\bibfnamefont{D.}~\bibnamefont{Vilone}}, \bibnamefont{and}
  \bibinfo{author}{\bibfnamefont{A.}~\bibnamefont{Vespignani}},
  \bibinfo{journal}{Europhys. Lett.} \textbf{\bibinfo{volume}{63}},
  \bibinfo{pages}{153} (\bibinfo{year}{2003});
\bibinfo{author}{\bibfnamefont{K.}~\bibnamefont{Suchecki}},
  \bibinfo{author}{\bibfnamefont{V.~M.} \bibnamefont{Egu{\'i}luz}},
  \bibnamefont{and}
  \bibinfo{author}{\bibfnamefont{M.}~\bibnamefont{San~Miguel}},
  \bibinfo{journal}{Europhys. Lett.} \textbf{\bibinfo{volume}{69}},
  \bibinfo{pages}{228} (\bibinfo{year}{2005}{\natexlab{b}}).

\bibitem[{\citenamefont{Dornic et~al.}(2001)\citenamefont{Dornic, Chat{\'e},
  Chave, and Hinrichsen}}]{col3}
\bibinfo{author}{\bibfnamefont{I.}~\bibnamefont{Dornic}},
  \bibinfo{author}{\bibfnamefont{H.}~\bibnamefont{Chat{\'e}}},
  \bibinfo{author}{\bibfnamefont{J.}~\bibnamefont{Chave}}, \bibnamefont{and}
  \bibinfo{author}{\bibfnamefont{H.}~\bibnamefont{Hinrichsen}},
  \bibinfo{journal}{Phys. Rev. Lett.} \textbf{\bibinfo{volume}{87}},
  \bibinfo{pages}{045701} (\bibinfo{year}{2001});
\bibinfo{author}{\bibfnamefont{L.}~\bibnamefont{Dall'Asta}} \bibnamefont{and}
  \bibinfo{author}{\bibfnamefont{C.}~\bibnamefont{Castellano}},
  \bibinfo{journal}{Europhys. Lett.} \textbf{\bibinfo{volume}{77}}
  (\bibinfo{year}{2007}).

\bibitem[{\citenamefont{Krapivsky}(1992)}]{krapivsky1992}
\bibinfo{author}{\bibfnamefont{P.}~\bibnamefont{Krapivsky}},
  \bibinfo{journal}{Phys. Rev. A} \textbf{\bibinfo{volume}{45}},
  \bibinfo{pages}{1067} (\bibinfo{year}{1992}).


\bibitem[{\citenamefont{Ravasz et~al.}(2004)\citenamefont{Ravasz, Szabo, and
  Szolnoki}}]{ravasz04}
\bibinfo{author}{\bibfnamefont{M.}~\bibnamefont{Ravasz}},
  \bibinfo{author}{\bibfnamefont{G.}~\bibnamefont{Szabo}}, \bibnamefont{and}
  \bibinfo{author}{\bibfnamefont{A.}~\bibnamefont{Szolnoki}},
  \bibinfo{journal}{Phys. Rev. E} \textbf{\bibinfo{volume}{70}},
  \bibinfo{eid}{012901} (\bibinfo{year}{2004}).

\bibitem[{\citenamefont{Molofsky et~al.}(1999)\citenamefont{Molofsky, Durrett,
  Dushoff, Griffeath, and Levin}}]{col4}
\bibinfo{author}{\bibfnamefont{J.}~\bibnamefont{Molofsky}},
  \bibinfo{author}{\bibfnamefont{R.}~\bibnamefont{Durrett}},
  \bibinfo{author}{\bibfnamefont{J.}~\bibnamefont{Dushoff}},
  \bibinfo{author}{\bibfnamefont{D.}~\bibnamefont{Griffeath}},
  \bibnamefont{and} \bibinfo{author}{\bibfnamefont{S.}~\bibnamefont{Levin}},
  \bibinfo{journal}{Theor. Pop. Biol.}
  \textbf{\bibinfo{volume}{55}}, \bibinfo{pages}{270} (\bibinfo{year}{1999});
\bibinfo{author}{\bibfnamefont{J.}~\bibnamefont{Chave}}, \bibinfo{journal}{
  Am. Nat.} \textbf{\bibinfo{volume}{157}}, \bibinfo{pages}{51}
  (\bibinfo{year}{2001});
\bibinfo{author}{\bibfnamefont{T.}~\bibnamefont{Zillio}},
  \bibinfo{author}{\bibfnamefont{I.}~\bibnamefont{Volkov}},
  \bibinfo{author}{\bibfnamefont{J.}~\bibnamefont{Banavar}},
  \bibinfo{author}{\bibfnamefont{S.}~\bibnamefont{Hubbell}}, \bibnamefont{and}
  \bibinfo{author}{\bibfnamefont{A.}~\bibnamefont{Maritan}},
  \bibinfo{journal}{Phys. Rev. Lett.} \textbf{\bibinfo{volume}{95}},
  \bibinfo{pages}{98101} (\bibinfo{year}{2005}).

\bibitem[{\citenamefont{Galam}(2005)}]{galam2005}
\bibinfo{author}{\bibfnamefont{S.}~\bibnamefont{Galam}},
  \bibinfo{journal}{Europhys. Lett.} \textbf{\bibinfo{volume}{20}},
  \bibinfo{pages}{705} (\bibinfo{year}{2005}).

\bibitem[{\citenamefont{Gunton et~al.}(1983)\citenamefont{Gunton, San~Miguel,
  and Sahni}}]{col5}
\bibinfo{author}{\bibfnamefont{J.~D.} \bibnamefont{Gunton}},
  \bibinfo{author}{\bibfnamefont{M.}~\bibnamefont{San~Miguel}},
  \bibnamefont{and} \bibinfo{author}{\bibfnamefont{P.~S.} \bibnamefont{Sahni}},
  in \emph{\bibinfo{booktitle}{Phase Transitions and Critical Phenomena}},
  edited by \bibinfo{editor}{\bibfnamefont{C.}~\bibnamefont{Domb}}
  \bibnamefont{and} \bibinfo{editor}{\bibfnamefont{J.}~\bibnamefont{Lebowitz}}
  (\bibinfo{publisher}{Academic Press}, \bibinfo{address}{London},
  \bibinfo{year}{1983}), vol.~\bibinfo{volume}{8}, pp.
  \bibinfo{pages}{269--446};
\bibinfo{author}{\bibfnamefont{C.}~\bibnamefont{Castellano}},
  \bibinfo{author}{\bibfnamefont{S.}~\bibnamefont{Fortunato}},
  \bibnamefont{and} \bibinfo{author}{\bibfnamefont{V.}~\bibnamefont{Loreto}}
  (\bibinfo{year}{2007}), \urlprefix\url{http://arxiv.org/abs/0710.3256}.

\bibitem[{\citenamefont{Cugliandolo and Kurchan}(1993)}]{cugliandolo1993}
\bibinfo{author}{\bibfnamefont{L.~F.} \bibnamefont{Cugliandolo}}
  \bibnamefont{and} \bibinfo{author}{\bibfnamefont{J.}~\bibnamefont{Kurchan}},
  \bibinfo{journal}{Phys. Rev. Lett.} \textbf{\bibinfo{volume}{71}},
  \bibinfo{pages}{173} (\bibinfo{year}{1993}).

\bibitem[{\citenamefont{Maslov et~al.}(2003)\citenamefont{Maslov, Sneppen, and
  Alon}}]{maslov2003}
\bibinfo{author}{\bibfnamefont{S.}~\bibnamefont{Maslov}},
  \bibinfo{author}{\bibfnamefont{K.}~\bibnamefont{Sneppen}}, \bibnamefont{and}
  \bibinfo{author}{\bibfnamefont{U.}~\bibnamefont{Alon}}, in
  \emph{\bibinfo{booktitle}{Handbook of graphs and networks. {F}rom the genoma
  to the internet}}, edited by
  \bibinfo{editor}{\bibfnamefont{S.}~\bibnamefont{Bornholdt}} \bibnamefont{and}
  \bibinfo{editor}{\bibfnamefont{H.}~\bibnamefont{Schuster}}
  (\bibinfo{publisher}{Wiley VCH and Co.}, \bibinfo{year}{2003}).

\bibitem[{\citenamefont{Helbing et~al.}(2000)\citenamefont{Helbing, Farkas, and
  Vicsek}}]{helbing}
\bibinfo{author}{\bibfnamefont{D.}~\bibnamefont{Helbing}},
  \bibinfo{author}{\bibfnamefont{I.}~\bibnamefont{Farkas}}, \bibnamefont{and}
  \bibinfo{author}{\bibfnamefont{T.}~\bibnamefont{Vicsek}},
  \bibinfo{journal}{Nature} \textbf{\bibinfo{volume}{407}},
  \bibinfo{pages}{487} (\bibinfo{year}{2000}).


\end{thebibliography}
\end{document}